\documentclass[12pt,preprint]{aastex}

\newcommand{\beq}{\begin{equation}}
\newcommand{\eeq}{\end{equation}}
\newcommand{\lsim}{\lesssim}
\newcommand{\gsim}{\gtrsim}

\begin{document}

\title{Luminosity Function of Morphologically Classified Galaxies in
the Sloan Digital Sky Survey}  

\author{Osamu Nakamura\altaffilmark{1},
Masataka Fukugita\altaffilmark{1,2},
Naoki Yasuda\altaffilmark{3},
Jon Loveday\altaffilmark{4},
Jon Brinkmann\altaffilmark{5}, 
Donald P. Schneider\altaffilmark{6},
Kazuhiro Shimasaku\altaffilmark{7}, 
Mark SubbaRao\altaffilmark{8}
}
\altaffiltext{1}{Institute for Cosmic Ray Research, University of Tokyo,
Kashiwa 2778582, Japan}
%%\email{osamu@icrr.u-tokyo.ac.jp}

\altaffiltext{2}{Institute for Advanced Study, Princeton, NJ 08540, U. S. A.}

\altaffiltext{3}{National Astronomical Observatory,
Mitaka, Tokyo 1818588, Japan}

\altaffiltext{4}{Astronomy Centre, University of Sussex, Brighton BN1 9QJ,
United Kingdom}

\altaffiltext{5}{Apache Point Observatory, Sunspot, NM 88349, U. S. A.}

\altaffiltext{6}{Department of Astronomy and Astrophysics, Pennsylvania 
State University, University Park, PA 16802, U. S. A.}

\altaffiltext{7}{Department of Astronomy, University of Tokyo, Tokyo 1130033,
Japan}

\altaffiltext{8}{Astronomy and Astrophysics Center, University of Chicago, 
Chicago, IL 60637, U. S. A.}

%\author{+ other SDSS members}
%\affil{XXX XXX XXX}
%\authoremail{pjep@pupgg.princeton.edu}
 
\begin{abstract}
The morphological dependence of the luminosity function is studied using
a sample containing approximately 
1500 bright galaxies classified into  Hubble types by visual inspections
for a homogeneous sample obtained from  
the Sloan Digital Sky Survey (SDSS)
northern equatorial stripes. 
Early-type galaxies are shown to have a characteristic
magnitude by 0.45 mag brighter than spiral galaxies in the $r^{\ast}$ band,
consistent with the `universal characteristic luminosity' in the
$B$ band. The shape of the luminosity function differs rather little
among different morphological types: we do not see any symptoms 
of the sharp decline in the faint end for
the luminosity function for early-type galaxies at least
2 mag fainter than the characteristic magnitude,
although the faint end behaviour shows a slight decline ($\alpha\lsim -1$) 
compared with the total 
sample. We also show that a rather flat faint end slope 
for early-type galaxies is not
due to an increasing mixture of the dwarf galaxies which have softer
cores. This means that there are numerous faint early-type galaxies with
highly concentrated cores.

\end{abstract}

\section{Introduction}

The origin of morphology of galaxies is a long-standing issue,
which could provide a key to discerning among models of the formation
of galaxies. How galaxy morphology changes 
as a function of the lookback time is perhaps
the prime approach to this problem, and the 
knowledge of the morphological dependence of the local luminosity function
at zero redshift is the baseline. 
One specific example of the issues 
is whether the luminosity function of elliptical galaxies 
obeys the Schechter-type function 
with a rather flat faint end (e.g., Marzke et al. 1994;
Kochanek et al. 2001), 
or the Gaussian function as inferred by  
Binggeli, Sandage \& Tammann (1988) and more recently 
by Bernardi et al. (2002a). If the latter is
correct, one would envisage  
galaxy morphology as a bulge-luminosity sequence (Dressler \& Sandage   
1983; Meisels and Ostriker 1984), which in turn reveals
a clue about the formation of elliptical galaxies and bulges.

There are also a number of uses of the morphology-dependent luminosity
function (MDLF). We mention only one example:
the frequency of gravitational lensing of quasar images is
approximately proportional to the luminosity density of early-type
galaxies rather than that of all galaxies (Fukugita \& Turner 1991). 
The uncertainty in
the MDLF is the largest source of error in predicting the frequency
of gravitational lenses, and thus in inferring the cosmological
constant from such analyses. 

The understanding of the MDLF is significantly
poorer than that of the luminosity function for galaxies in general, 
which has undergone
substantial progress in the latest years (Folkes et al. 1999;
Blanton et al. 2001) by virtue of large galaxy samples. 
The traditional way to obtain MDLF is to use morphological 
classification based 
on visual inspections of the images (Binggeli et al. 1988;
Loveday et al. 1992; Marzke et al. 1994; 1998; Kochanek et al. 2001).
Some modern  studies attempt to 
use spectroscopic features to classify galaxies into 
morphological types (Bromley et al. 1998; %Loveday, Tresse \& Maddox 1999;
Folkes et al. 1999), which makes 
it possible to analyze large samples. 
Although a general correlation is known between
spectroscopic and Hubble morphologies, the samples derived from the two
methods are considerably different. In particular, the classification
using spectroscopic features or colours is sensitive to small star
formation activities now or in the near past in early-type galaxies, 
whilst Hubble
morphology is insensitive to this process.
The problem of automated classification always lies in the
difficulty in finding quantitative measures that strongly
correlate with the Hubble sequence based on visual inspections.
%(we call this Hubble morphology). 
In this paper we derive the MDLF based on visual classifications using a
homogeneous bright galaxy sample from Sloan Digital Sky Survey (SDSS;
York et al. 2000). The sample we use in this paper is small, 
but it is based on a homogeneous morphological classification with
accurate photometry.  

The SDSS conducts both photometric
(Gunn et al. 1998; Hogg et al. 2001; Pier et al. 2002)
and spectroscopic surveys, and is producing a homogeneous 
data set, which is suitable to studies of galaxy statistics. 
The initial survey observations were made in the northern and  
southern equatorial stripes, and produced a galaxy catalogue to $r^{\ast}=22.5$
mag in five colour bands (Fukugita et al. 1996) with a photometric
calibration using a new standard star network observed at 
USNO (Smith et al. 2002). 
Spectroscopic follow-up is made to 17.8 mag with accurately defined 
criteria for target selection (Strauss et al. 2002). 

Our study is limited to bright galaxies with $r^{\ast}\leq 15.9$ mag after 
Galactic extinction correction, since visual classifications sometimes
cannot be made confidently beyond this magnitude with the SDSS imaging
data. We have classified all galaxies satisfying this magnitude criterion
in the northern equatorial stripe. The total
number of galaxies in our sample is 1875, of which 1600 have 
spectroscopic information.

The dominant part of the data we used are
already published as an {\it Early Data Release} (EDR) (Stoughton et al.
2002). Our present work uses primarily the EDR but supplemented by
observations which are not included in EDR to make the sample 
as complete as possible. Photometry of galaxies in this region
is discussed in a galaxy number count paper of 
Yasuda et al. (2001), and the luminosity function is derived by
Blanton et al. (2001), which also discuss spectroscopic details.

\section{The sample and the morphology classification}

The region of the sky we consider is the northern equatorial stripe 
(SDSS photometry run numbers 752 and 756)
for $145.15^\circ\leq \alpha({\rm J2000})\leq 235.97^\circ$ and 
$|\delta({\rm J2000})|\leq 1.27
^\circ$,
% The region we consider is a rectangular field circumscribed by RA and dec, 
which is included 
in the EDR sample. The total area is 229.7 square deg. 
We apply Galactic 
extinction correction using the extinction map of Schlegel, Finkbeiner
\& Davis (1998) 
assuming $R_{r^{\ast}}=A_{r^{\ast}}/E(B-V)=2.75$,
and select galaxies with the Petrosian magnitude 
${r^{\ast}}_P\leq 15.9$ after the correction in the automatically-generated 
photometric catalogue (Stoughton et al. 2002). 
We use the extinction-corrected Petrosian magnitude throughout this paper.

The photometric catalogue yields 2418 galaxy candidates with 
${r^{\ast}}_P\leq 15.9$ if we follow the criteria given in Strauss et al. (2002). 
This sample still contains number of double stars and
shredded galaxies due to deblending failures,
which cannot be rejected by the automated algorithm.
We obtain after visual inspection of all galaxy candidates 1875 galaxies, 
of which 1600 (85\%) are 
included in the spectroscopic sample. 
%(The remainders, 2418$-$1875, are mostly
%double stars and shredded galaxies due to deblending failures).
Spectroscopy was made using 50 plugged plates with additional 41 plates that
are centred in the neighbouring stripes. These plates cover 228.1 square 
deg.
The confidence level for the redshift determination is mostly over 99\%, but
9 galaxies are given low ($<85$\%) confidence, which we omit from 
our sample. We also drop 38 galaxies which either contain multiple
galaxies or have poor photometry due to deblending failures. 
This leaves 1553
galaxies. We note that there are some galaxies which are dropped in the
primary galaxy selection in the photometric catalogue 
(Yasuda et al. 2001; Strauss et al. 2002) due
to saturation flags caused by nearby bright stars or to other reasons. We
estimate that we have probably missed about $\approx$88 galaxies 
in our field from the rate of missed galaxies 
given in Yasuda et al. So the overall
sample completeness is estimated to be 79.5\%. For more detailed discussion for
the spectroscopic sample, see Blanton et al. (2001).

All galaxies in our sample (1875) 
are classified into 7 morphological classes, 
$T=0$ (corresponding to E in the Hubble type), 
1 (S0), 2 (Sa), 3 (Sb), 4 (Sc), 5 (Sd), and 6 (Im). 
Morphology classification is carried out by two of us 
(MF and ON) using the $g^{\ast}$ band image of each galaxy displayed 
on the SAOimage viewer, according
to {\it Hubble Atlas of Galaxies} (Sandage 1961).
%\footnote{After completion 
%of this work we added two more classifiers and made our final catalogue
%of morphologically classified galaxies. Since
%the catalogue we use here and the one newly generated
%differ only in fine details, and the use of the new catalogue
%does not modify the result, we present
%in this paper the result based on the catalogue by two classifiers.}. 
We also refer to morphological types given by the 
{\it Third Reference Catalogue of
Bright Galaxies} (de Vaucouleurs et al. 1991; RC3), so that
our classification closely matches to the traditional scheme,
although the RC3 classification, which is based on the photographic
material, is occasionally incorrect when galaxies are viewed with 
the CCD image, with which we can look at the image with different
levels of brightness and contrast. 
We give an index of $-1$ when we cannot assign a morphological type.
The classification by the two independent visual inspections
agrees to within $\Delta T\leq 1.5$ 
for most galaxies and a mean (0.5 step in $T$) is taken for our 
final classification.

We reclassify galaxies into three groups of $T$,
$0\le T\le 1.0$ (E-S0), $1.5\le T\le 3$ (S0/a-Sb), and
$3.5 \le T\le 5$ (Sbc-Sd).
The morphological distributions of the galaxies in the different
samples are given in Table 1.
The ratio of E-S0:S0/a-Sb:Sbc-Sd:Im $\simeq$ 0.40:0.34:0.24:0.02.
This is a somewhat larger fraction of E-S0 compared to the value 
usually adopted due to our use of
$r^{\ast}$ colour as the prime passband. For the same reason the fraction
of Im galaxies is smaller by a factor of 2-3 than that from 
$B$ selected samples. In this work we do not divide the morphology
into further detailed classes considering the uncertainty in visual
classification, especially between E and S0 for fainter galaxies.  
We consider $T>5$ galaxies separately, since the 
spectroscopic target selection is biased against
low surface brightness galaxies,  and the relatively low quality
of photometry for this class of galaxies makes the incompleteness 
significant; the completeness fraction of Im galaxies as read from Table 1 
is only 54\%, which is compared
with $\approx$83\% for other classes of galaxies. 
Along with a small Im fraction, 
our sample for an appropriate redshift range 
is too small to derive a reliable  MDLF for Im galaxies.   
%The catalogue of morphologically classified galaxies will be published 
%elsewhere (Fukugita et al., unpublished).

\section{Morphology-Dependent Luminosity Functions}

We show in Figure 1 the differential number counts of galaxies.
The slope of the counts is slightly steeper than that of the
Euclidean value, and is in agreement with previous studies 
(Yasuda et al. 2001). The counts of spectroscopic galaxies
(indicated by the dashed curve) follow those of the photometric 
sample within one sigma 
of Poisson statistics, so that the completeness correction for
the sample of the present paper does not depend on brightness. 
%This ensures that the sample incompleteness is homogeneous 
%with respect to brightness of galaxies.

We use the recession velocity with respect to the Galactic Standard of Rest
according to RC3.  
We select galaxies in the redshift range 
3000 km s$^{-1}<cz< 36000$ km s$^{-1}$. The lower cutoff
is imposed to avoid large effects from peculiar velocity flow, and
the upper cutoff is practically the limit of our sample. 
We further impose a cut on apparent
magnitude as $r^{\ast}\geq 13.2$ mag, since very bright galaxies are often 
dropped from spectroscopic targets. 
These selections exclude 71 galaxies from our sample,
leaving 1482 galaxies used to estimate the MDLF. The redshift distributions
of our galaxy sample are shown in Figure 2, where the curves show
expectations for a homogeneous universe with the MDLF derived 
in this paper.

We compute MDLFs for the samples 
with three methods: maximum-likelihood (ML) (Sandage,
Tammann \& Yahil 1979), 
step-wise maximum-likelihood (SWML) (Efstathiou, Ellis \& Peterson 1988)
and the $V_{\rm max}$ method. We take the step of luminosity to be
0.25 mag for SWML.
We adopt $\Omega=0.3$ and $\lambda=0.7$ for cosmology, although 
the maximum redshift of our sample is $z=0.12$ and 
the results hardly depend on the cosmological parameters. 
The $K$ correction is 
taken from Fukugita, Shimasaku \& Ichikawa (1995) with an interpolation
with respect to $g^{\ast}-r^{\ast}$ colour for each galaxy.

The results from the first two methods, ML and SWML, show
a good agreement, but those from the $V_{\rm max}$ method
differ from the former two in the faint end. This is a
well-known effect generally ascribed to inhomogeneous galaxy 
distributions in the redshift space, as are visible in Figure 2.
In Figure 3 we
present the MDLF from ML and SWML in the $r^{\ast}$ passband, 
together with the absolute magnitude distribution  
of galaxies used in the analysis. The ML estimate assumes 
the Schechter function 

\begin{equation}
\phi(L)dL=\phi^*\left({L\over L_*}\right)^\alpha 
\exp\left[-\left({L\over L_*}\right)\right]{dL\over L_*} \ ,
\label{eq:schechter}
\end{equation}
and the derived parameters are given in Table 2, where we take
the Hubble constant $h=H_0/100$ km s$^{-1}$Mpc$^{-1}$. 
Only a crude estimate (with ML) is presented 
for the luminosity function for Im galaxies, since our sample is
too small. 
We also present the results for the total sample, which includes
not only galaxies with $T=0-6$, but also those could not be classified
($T=-1$). This is a bright-galaxy version of the analysis
given by Blanton et al. (2001).

In this table we also give the luminosity densities obtained by
integrating (\ref{eq:schechter}) over $L=0$ to $\infty$. 
The contours of  
one and two standard deviation errors calculated from the likelihood
functions are shown in $\alpha-M^*$ plane in 
Figure 4. We have also carried out a jack-knife error estimate,
by dividing the sample into ten RA bins (width of $\sim$1.2 hr), 
in order to study the 
effect of the sample variance. The best-fit values for the subsamples
all fall within the one-sigma ellipse given above, and the
variance estimated from the jack-knife method is smaller than the
error we quoted. So we adopt one-sigma of the fit for our final
error estimate. 

We then assign an additional 0.05 mag error from the calibration of
photometry (added in quadratures). 
For more discussion about errors and selection
effects, see Blanton et al. (2001). The errors expected from a number
of items seen in their analysis are significantly smaller than 
the statistical error we are concerned with here.

We determine the normalization $\phi^*$ of the MDLF following the method of
Efstathiou et al. (1988) for each sample of morphologically-classified 
galaxies. We adopt the region of $M_{r^{\ast}}$ where the sample contains
sufficient number of galaxies, dropping too bright ($M_{r^{\ast}}<M_{r^{\ast}}^*-2$)
and faint ($M_{r^{\ast}}>M_{r^{\ast}}^*+1$) galaxies and those with high redshifts
to avoid strong shot noise effects. We choose the redshift range 
to be $0.01\le z\le 0.075$, for which the selection function for the
total sample is $\gsim 0.14$. 
The numbers of galaxies used to determine $\phi^*$ are given in
Table 1 above. In Table~2 we give jack-knife errors for $\phi^*$. 
The normalization significantly varies depending on the cutoff of the redshift 
range, reflecting the presence of large-scale structure,
such as a clump seen between $z=0.07$ and 0.08 in Figure~2.
The variation of the normalization by varying the upper cutoff between 
0.07 and 0.08 is comparable to the jack-knife error we quoted.
The normalizations (and errors) are then corrected for the sample 
incompleteness derived in Table 2 by comparing the spectroscopic sample
with good-quality photometry and redshift determinations [(d) in Table 1]
to the photometric sample [(a) in Table 1]. A small difference of areas
covered by photometric and spectroscopic surveys is also taken into account. 
Furthermore, an extra correction factor
of (1875+88)/1875=1.05 is multiplied to correct for the 
incompleteness of the photometric catalogue as discussed in section 2.
% which is estimated
% from the comparison of the machine-selected with the visually-selected
% sample in Yasuda et al. (2001) for $r^{\ast}<16$ before the 
% extinction correction. Our final results are denoted by $\phi_c^*$.
% We adopt jack-knife errors for $\phi^*$ and $\phi_c^*$ by dividing the
% area into 10 according to RA.

We can see following features in our luminosity functions:

(i) The characteristic luminosity and the faint end slope of the total 
sample are consistent with the parameters derived by 
Blanton et al. (2001) within 1$-$1.3 sigma. The normalization, however,
is significantly lower, corresponding to by 30\% in the
luminosity density, than that of Blanton et al. This is 
ascribed to the local deficit of galaxies in the northern 
equatorial stripe seen
for $r^{\ast}<16$ mag, and is ascribed to large-scale structure, as discussed 
in Yasuda et al. (2001). 
We confirmed that the normalization rapidly
approaches that of Blanton
et al. when we take the limiting magnitude fainter; with $r^{\ast}<16.5$ mag,
the luminosity density agrees with that of Blanton et al. within 10\%.

(ii) The characteristic luminosity of early-type galaxies is 
more luminous than that of later-type galaxies by about 0.45 mag. 
This is consistent with the `universal characteristic luminosity'
known for the $B$ band (Tammann, Yahil \& Sandage 1979), because
we expect that $B-r^{\ast}$ colour differs by 0.4 mag between E and Sb
(Fukugita et al. 1995). This implies that the 
universal characteristic luminosity in the $B$ band is 
an accidental effect.

(iii) The shape of the luminosity function of early-type galaxies
is not much different from that of late-type galaxies, although
some trend is seen that the number of early-type galaxies  
slightly declines ($\alpha\lsim -1$) towards the faint end. 
This conclusion agrees with Marzke et al. (1994) for the $B$ band, 
and Kochanek et al. (2001) for the $K$ band,
but does not agree with Loveday et al. (1992), 
which show an appreciable decline towards the faint end
(see Zucca, Pozzetti \& Zamorani 1994, which ascribe Loveday et al.'s
result to a sample incompleteness). 
In particular, we do not see a sharp decline of the luminosity 
function, as inferred in Binggeli et al. (1988) and Bernardi et al. (2002a). 
The latter authors fit the luminosity function of early-type
galaxies selected with photometric and spectroscopic parameters
(Bernardi et al. 2002b)\footnote{These authors adopted selection criteria
based on the concentration index ($C<0.4$; see below), the PCA spectral
classification index for early-type galaxies, and 
the de Vaucouleurs- versus exponential-likelihood parameters which are produced
from the photometric pipeline. Note that the last parameter correlates
very weakly with visual morphology; see Shimasaku et al. 2001.}   
to a Gaussian function with a peak at 
$M_{r^{\ast}}=-20.38$ mag ($h=1$): their data go beyond the peak
only slightly, and the turn-over is not conclusive. 
Our luminosity function, which goes down to $-$18.75 mag, 
does not show any turnover to this magnitude.

(iv) The luminosity function of late-type spirals (Sbc-Sd) does not exhibit
an increase ($\alpha\gsim -1$) towards the faint end. 
Our late-type spiral galaxy sample shows an even faster decline compared to
that of early-type spiral galaxies.
We found that the luminosity
function derived from $V_{\rm max}$ shows a somewhat
faster increase ($\alpha=-1.16$)
compared with those for other types, but this trend is not visible with the
MDLF from the ML or SWML methods. We ascribe this larger $\alpha$ from
the $V_{\rm max}$ method
to a local
effect of the galaxy distribution, as we mention below. 
In any case %the difference of the slope between
%early-and late-type spirals is not large %($|\delta\alpha|\lsim0.2$), and 
the steepening of the faint end slope does not occur up to 
the Im type. 
This might appear to contrast with the conventional belief that
late-type galaxies have a steep slope. This is due to our
exclusion of very late galaxies ($T>5$), and is consistent with
Marzke et al. (1994), who found that only the Im luminosity function 
shows a steep faint end slope.

(vi) The Im type luminosity function shows a steep faint end slope,
$\alpha\sim -1.9$, consistent with Marzke et al.

It may be worth commenting that the absolute magnitude distributions
of early- and late-type spiral galaxies shown in Figure 3 appear
to indicate steeper faint end slope for the latter. This is in fact
what we have obtained when we use the $V_{\rm max}$ method. This
reflects the effect seen in Figure 2 that the morphological
composition appears to change as a redshift [i.e., the frequency 
of late-type spirals is high in nearby ($z<0.05$) sample].
This effect disappears when we use the likelihood method to
calculate the luminosity function {\it under the assumption that
it is universal}. 

For practical uses of the MDLF presented here, the normalization should
be multiplied by a factor of 1.29 to correct for the local deficit of
galaxies in the northern equatorial stripe in brighter magnitudes.

\section{Morphological classification with the concentration index}

The luminosity function of early-type galaxies we derived 
does not show a conspicuous
decline towards the faint end. One may suspect that 
that our E and S0 sample may contain increasingly more 
dwarf ellipticals and spheroidals, which are not separated
from their giant counterparts in visual classifications,
towards the faint end, and therefore the luminosity function of
giant elliptical galaxies might actually decline.

This point may be studied by using a concentration index, 
since early-type dwarfs usually
have galaxy cores significantly softer than those of giant 
elliptical galaxies (e.g., Kormendy 1986).  
We define the (inverse) concentration index by the ratio of the two
Petrosian radii $C=r_{50}/r_{90}$ measured
in the $r^{\ast}$ band, where $r_{50}$ and $r_{90}$ are radii which correspond
to the apertures that include 50\% and 90\% of the Petrosian flux.
In Figure 5 we plot $C$ as a function of absolute magnitudes for E and S0
galaxies. The plot shows that there is no evident trend that fainter early-type
galaxies have softer cores; most of the data points fall below 
$C<0.34$, which is a typical value that divides early and late types, 
down to $-$19 mag.

Shimasaku et al. (2001) report that this $C$
parameter shows the strongest correlation with visually-classified 
morphology among simple photometrically-defined 
parameters (see also, Doi, Fukugita
\& Okamura 1993; Abraham et al. 1994; Blanton et al. 2001;
Strateva et al. 2001; Bernardi et al. 2002b). 
We thus separate morphologies into early and late types 
according as $C<0.35$ or $C>0.35$, which corresponds to the
devision at S0/a. The early-type galaxy sample (706 galaxies)
thus defined shows a 82\% completeness and is contaminated by
late-type galaxies by 18\% when we take the visually-classified
sample as the reference. The late-type sample (713 galaxies)
also shows a 82\% completeness and a 18 \% contamination from
the opposite sample. This choice of $C$ minimizes the contamination
of the opposite morphologies either way.
The analysis is similar to what
was already presented by Blanton et al. (2001), with the difference that
they have used $C=0.43$ (which corresponds to Sb for bright galaxies) 
to divide the early- and late-type galaxy samples. 

Figure 6 shows the MDLF separated according to this $C$ index.
The parameters of the Schechter function from the ML analysis 
are given in Table 3 above. The use of a different division at
$C=0.34$, which is about the division at S0 galaxies, changes the
MDLF only slightly. 
The feature of the luminosity functions is
similar to what are derived from the visually-classified sample.
The MDLF for early-type galaxies shows a characteristic 
luminosity brighter than that for late-types, and has a slightly
declining faint-end shape 
while late-type galaxies show a flat faint
end.  
No sharp decline of the luminosity function is visible at least two
mag fainter than $M^*$, or at least 2 mag fainter than the
peak inferred by Bernardi et al. (2002a). 

Our result  means that it is unlikely that the
visually-classified sample is dominated by dwarf ellipticals
that have soft cores in faint magnitudes.  
The luminosity function of late-type galaxies
also shares the features of the visually-classified late-type galaxy sample.
This analysis shows that the MDLFs using the concentration index 
are similar to those with visually-classified samples, although a
somewhat smaller difference of characteristic luminosities of the
two types represents the $\sim$20\% contamination from
the opposite types.

\section{Conclusions}

Our sample is small and we may not be able to extract 
quantitatively robust parameters, yet we obtain a number of
useful conclusions. The most important feature with our analysis is 
that we have used a homogeneous photometric catalogue with sharply defined
selection criteria and a homogeneously
morphologically-classified 
sample based on Hubble morphology of galaxies, rather than 
a sample classified by
indicators using spectroscopic
features or colours, which are sensitive to small star formation
activities in the present or the near past.  

The first conclusion we have obtained is that the shape of the
MDLF does not depend
too strongly on the Hubble types. The characteristic luminosity of
elliptical and S0 galaxies is brighter than that of spiral galaxies 
in the $r^{\ast}$ band.
The amount of the difference in brightness is consistent with 
universal characteristic luminosity in the $B$ band, which was
found by Tammann et al. (1979).   
The MDLF of early-type galaxies somewhat
declines in the faint end, but does not exhibit a sharp decline, and
this is not due to an increasing mixture of dwarf
galaxies at least in the magnitude range we are concerned with.
The conclusion is unchanged if we use
the concentration index as a classifier of early-type galaxies.
This indicates that there are many intrinsically faint 
elliptical galaxies, whose luminosities are fainter than those 
of bulges in spiral galaxies. 
The existence of numerous early-type galaxies with a hard core 
at small luminosities indicates that morphology is unlikely
to be a sequence of the bulge luminosity as advocated by 
Dressler \& Sandage (1983), and by Meisels and Ostriker (1984).
Our conclusion also justifies the calculation of the strong gravitational
lensing frequency of quasars using the standard Schechter function
%with the normalization simply scaled to the numbers of early-type
%galaxies, 
without introducing a cutoff in the luminosity function, which
would affect the frequency of sub-arcsecond lensing.

\vspace{10pt}

Funding for the creation and distribution of the SDSS Archive has been provided
by the Alfred P. Sloan Foundation, the Participating Institutions, the National
Aeronautics and Space Administration, the National Science Foundation, the U.S.
Department of Energy, the Japanese Monbukagakusho, and the Max Planck
Society. The SDSS Web site is http://www.sdss.org/. 
The SDSS is managed by the Astrophysical Research Consortium (ARC) for the
Participating Institutions. The Participating Institutions are The University of
Chicago, Fermilab, the Institute for Advanced Study, the Japan Participation
Group, The Johns Hopkins University, Los Alamos National Laboratory, the
Max-Planck-Institute for Astronomy (MPIA), the Max-Planck-Institute for
Astrophysics (MPA), New Mexico State University, University of Pittsburgh,
Princeton University, the United States Naval Observatory, and the University of
Washington. We would like to thank Sadanori Okamura for useful comments.
MF is supported in part by the Grant in Aid of the
Japanese Ministry of Education.

\clearpage

%figure 1
%%\vspace{0.5cm}
\begin{figure}
%%\psbox[width=8.5cm,aspect=1.0]{fig_magdist.eps}
%%\plotone{fig_magdist.eps}
\plotone{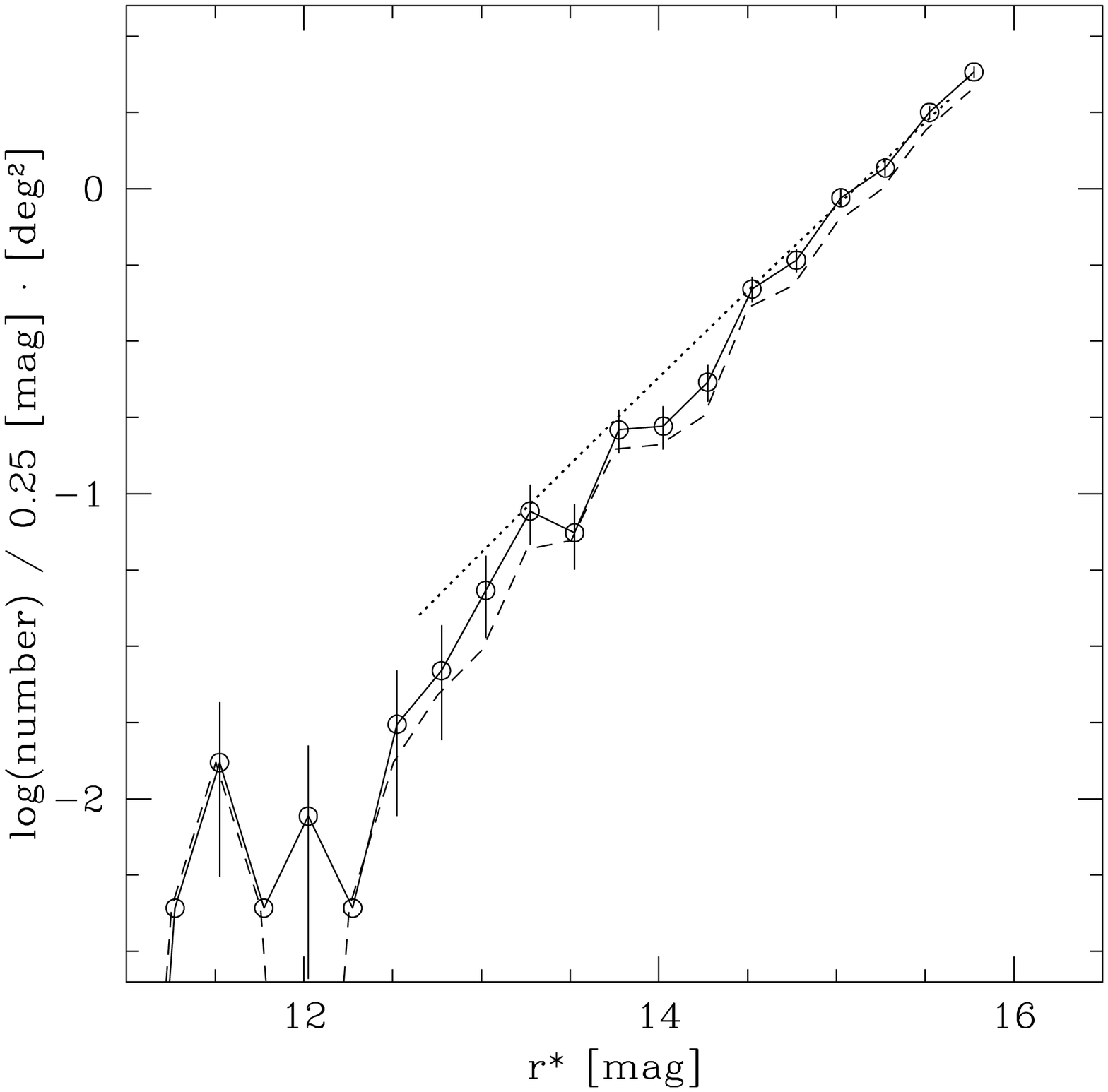}
\caption{
Differential galaxy number counts in the $r^{\ast}$ band 
as a function of magnitude. The dashed curve
is the spectroscopic sample, which is compared with the photometric sample
represented by the solid curve. The dotted curve is the prediction from
the luminosity function obtained in this paper, showing a deficiency of
galaxies in bright magnitudes for the northern equatorial stripe.}
\end{figure}%

%figure 2
\begin{figure}
%%\vspace{0.5cm}
%%\psbox[width=8.5cm,aspect=1.0]{fig_zdist.eps}
%%\plotone{fig_zdist.eps}
\plotone{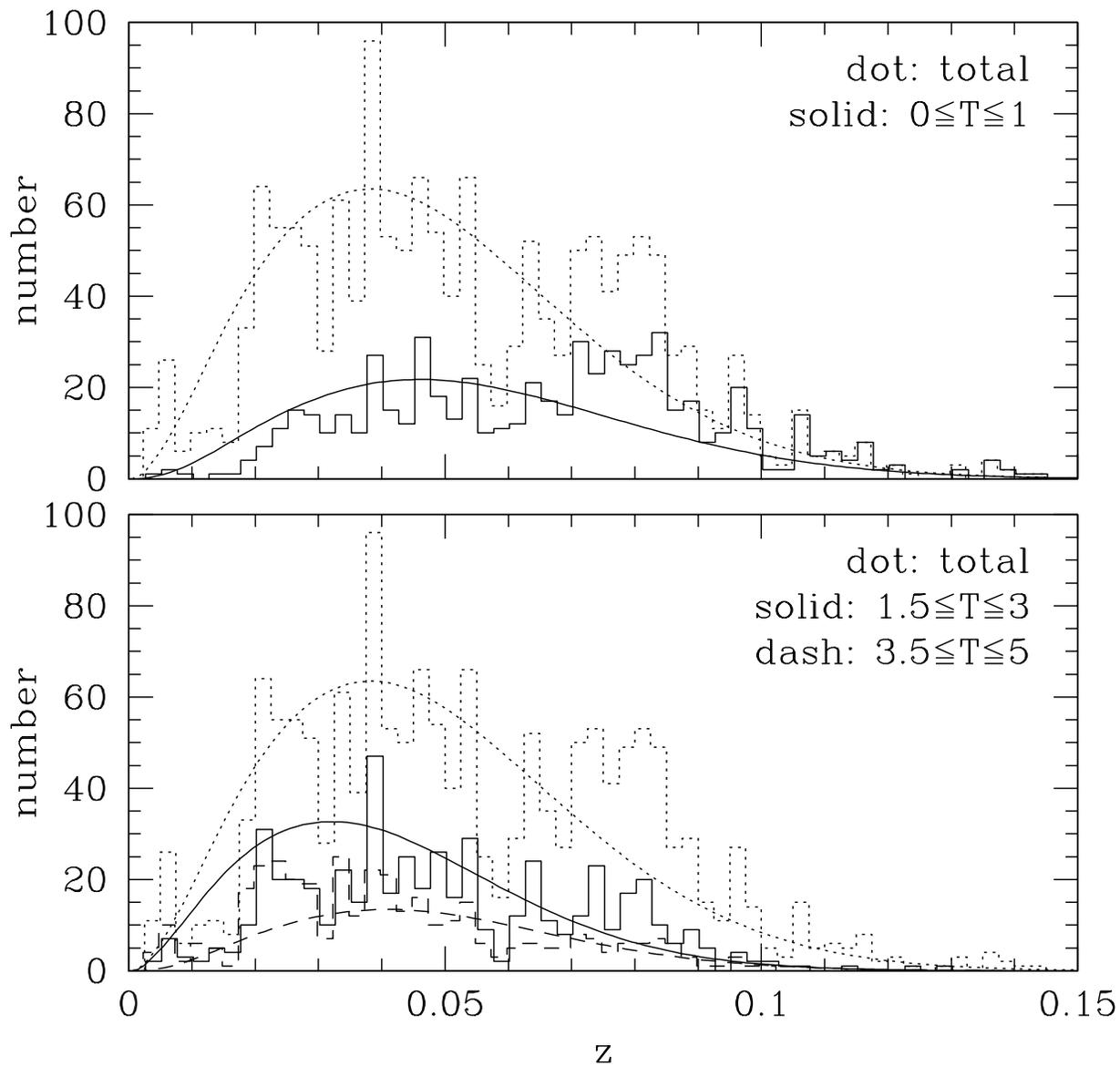}
\caption{
The redshift distribution of galaxies, compared with the
predictions for the homogeneous universe calculated with the 
MDLF we obtain in this paper. The upper panel is for
early-type (E and S0) galaxies, and the lower panel for later type 
(S0/a-Sb; Sbc-Sd) galaxies.
The dashed curves show the total sample.}
\end{figure}%

%figure 3
\begin{figure}
%%\vspace{0.5cm}
%%\psbox[width=8.5cm,aspect=1.0]{fig_mdlf.eps}
%%\plotone{fig_mdlf.eps}
\plotone{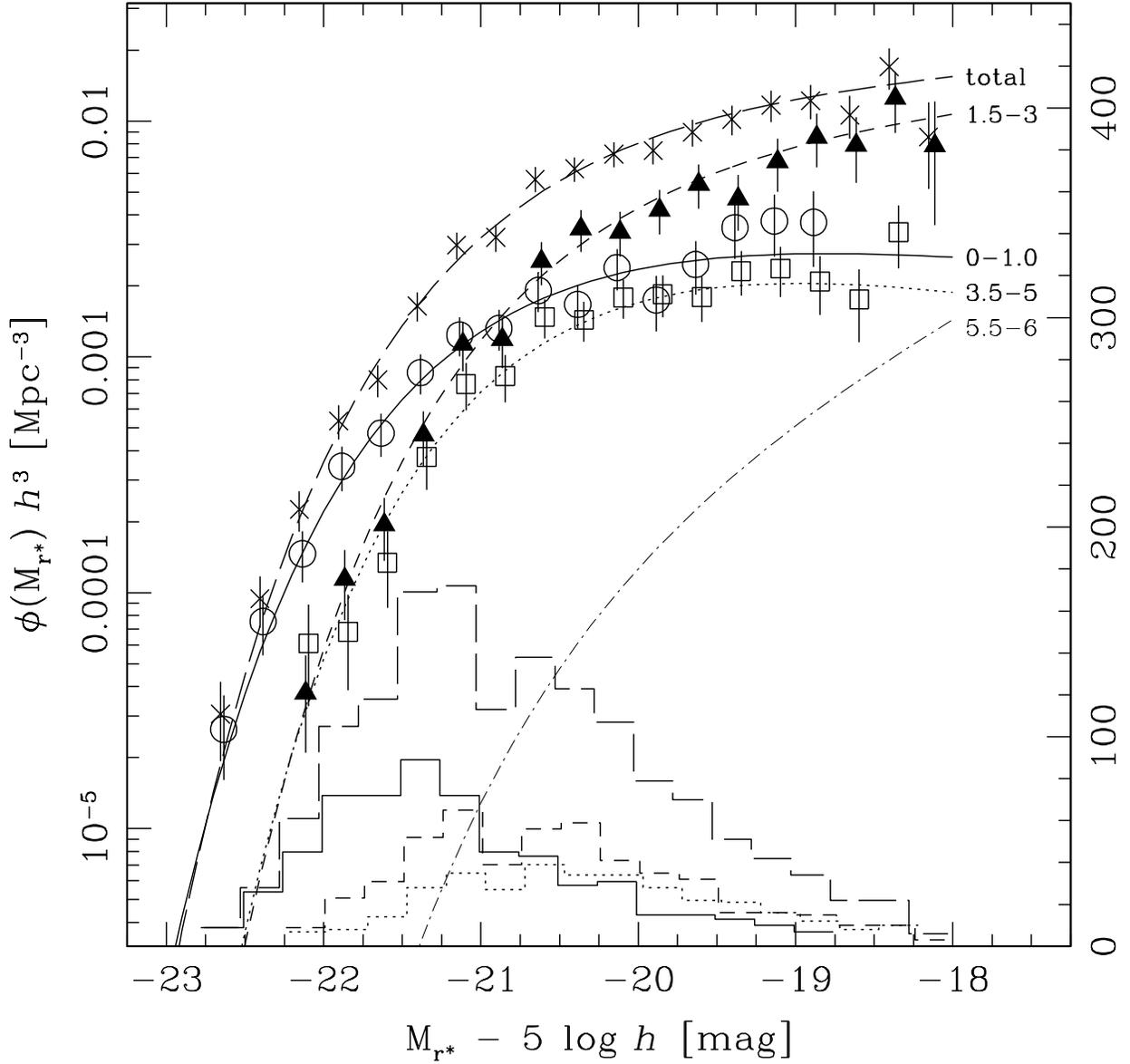}
\caption{
MDLF in the $r^{\ast}$ band for three types, E-S0, S0a-Sb, 
and Sbc-Sd from visual
classifications. The SWML results are represented by data points (open
circle for E-S0; solid triangle for early spiral galaxies;
open square for late-type spiral galaxies), and the ML fits are shown
by solid, short dashed and dotted curves, respectively. The ML
estimate for Im galaxies is represented by dash-dotted curve.
The luminosity function for the total sample is also
plotted for comparison, represented by crosses and  a long-dashed curve.
The histograms are the actual numbers of galaxies for the three 
types and the total sample used in this analysis.}
\end{figure}%

%%figure 4
\begin{figure}
%%\vspace{0.5cm}
%%\psbox[width=8.5cm,aspect=1.0]{fig_mdlf_err.eps}
%%\plotone{fig_mdlf_err.eps}
\plotone{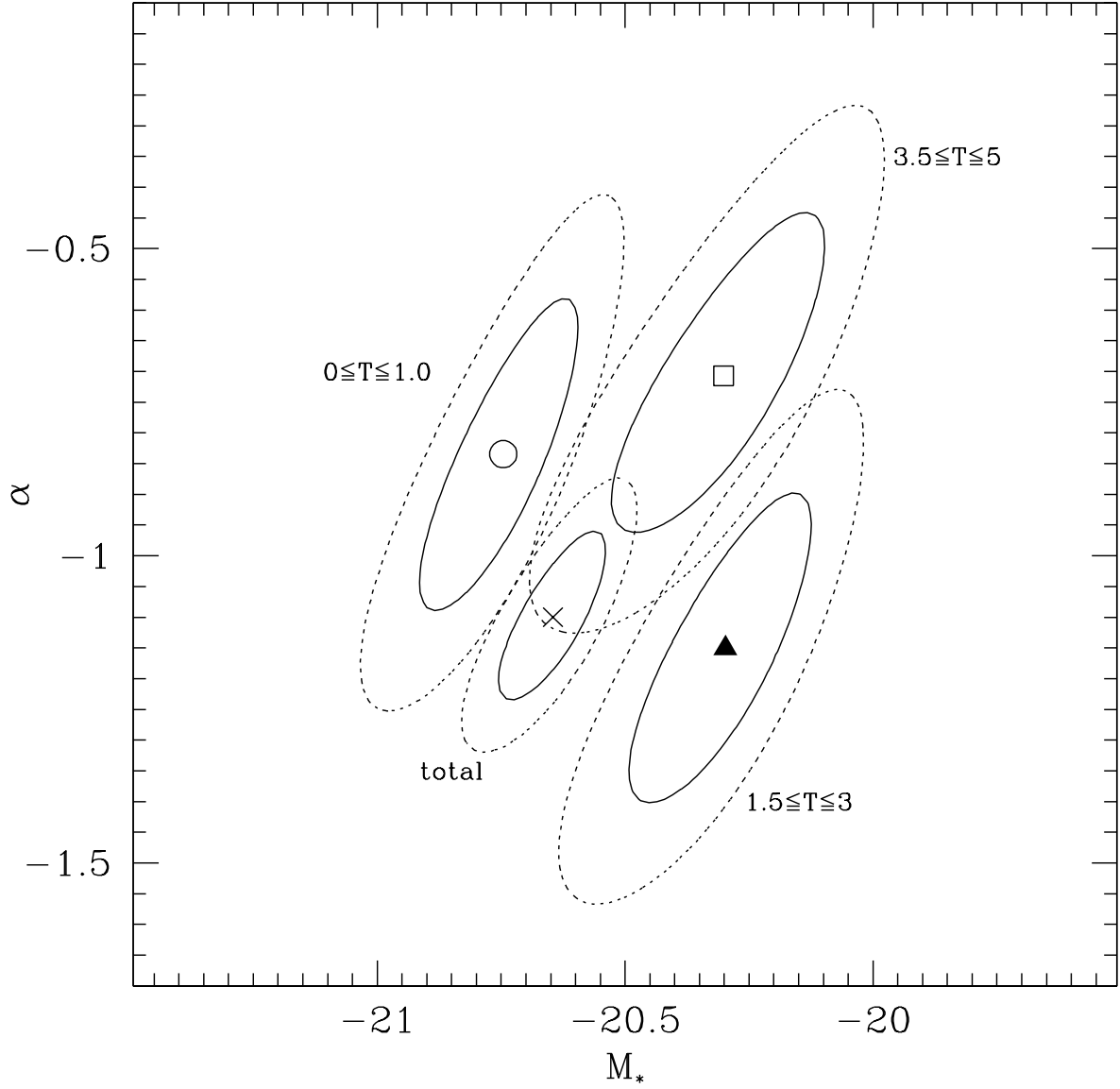}
\caption{
Error contours (solid curve for 1 sigma, dotted curve for
2 sigma) of the parameters of the
Schechter functions for the ML fits given in Figure 3.
The unit of the abscissa assumes $h=1$.}
\vspace{0.5cm}
\end{figure}%

%figure 5
\begin{figure}
%%\vspace{0.5cm}
%%\psbox[width=8.5cm,aspect=1.0]{fig_cin.eps}
%%\plotone{fig_cin.eps}
\plotone{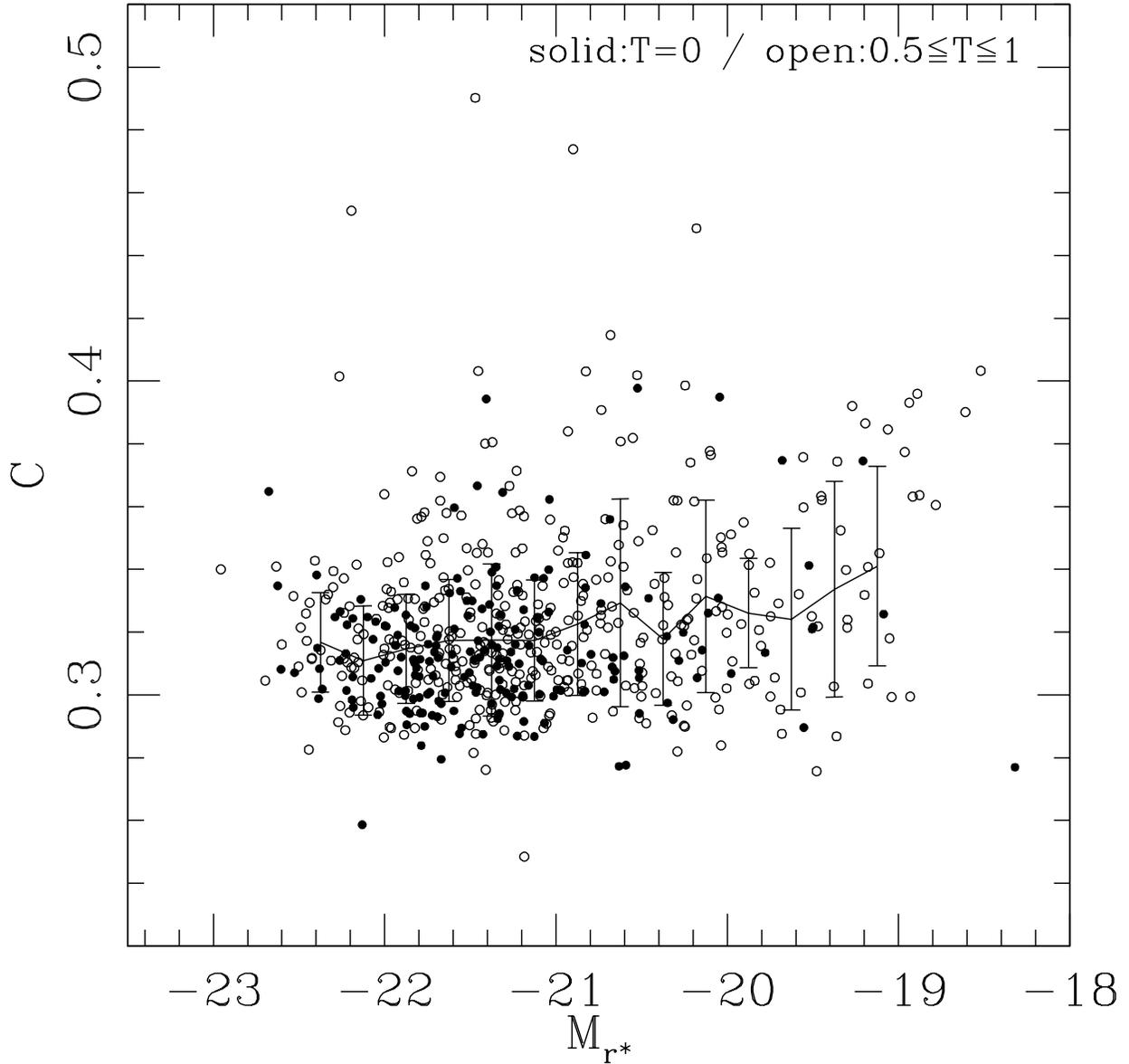}
\caption{
Concentration indices of early-type galaxies plotted as a 
function of absolute magnitudes. Solid points are galaxies classified
as E and open points are other early-type galaxies (E/S0 or S0). 
The average of $C$ as a function of luminosity is also given.
The unit of the abscissa assumes $h=1$.}
\end{figure}%

%figure 6
\begin{figure}
%%\vspace{0.5cm}
%%\psbox[width=8.5cm,aspect=1.0]{fig_mdlf_cin.eps}
%%\plotone{fig_mdlf_cin.eps}
\plotone{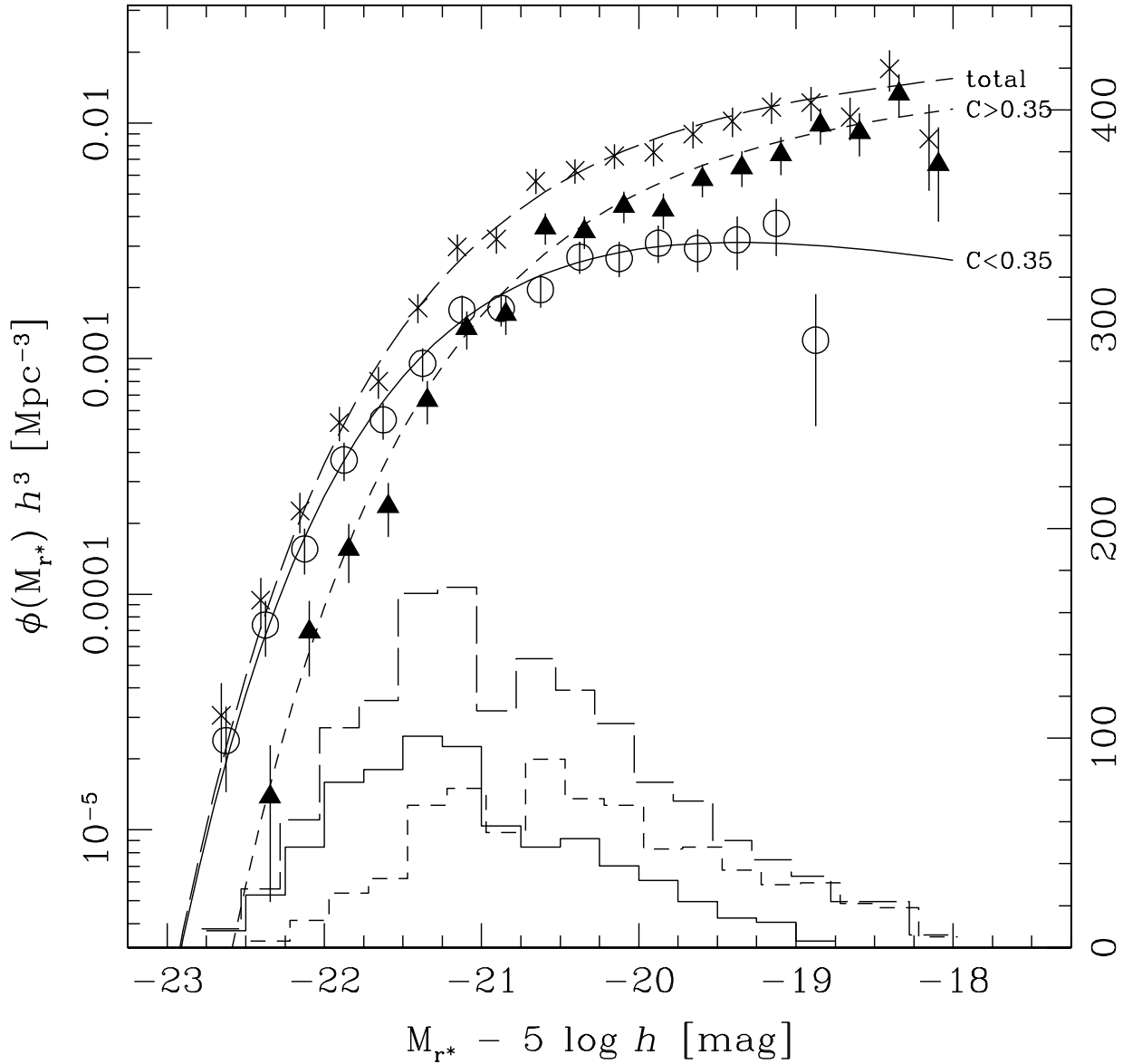}
\caption{
MDLF for early and late-type galaxies classified using the 
concentration index.  
The SWML results are represented by data points (open
circle for early-type with $C<0.35$; solid triangle for 
late type $C>0.35$), and the ML fits are shown
by solid and short dashed curves, respectively. 
The luminosity function for the total sample is 
plotted for comparison, represented by crosses and a long-dashed curve.
The histograms are the numbers of galaxies  
used in this analysis.}
\end{figure}%

\clearpage

%table 1
\begin{table*}
\begin{center}
\caption{Morphologically classified sample}
\begin{tabular}{lcccccc}
\tableline \tableline
  &  $0\le T\le 1$  & $1< T\le 3$  &
$3< T\le 5$ & $5< T\le 6$ & $T=-1$ & total \\
   & (E \& S0) & (S0/a-Sb) & (Sbc-Sd) & (Im) & (unclass.) & \\
\tableline
(a) photometric sample              & 740 & 630  & 444 & 35 & 26 & 1875   \\
(b) spectroscopic sample            & 630 & 545  & 381 & 23 & 21 & 1600   \\
(c) sample with good photometry     & 617 & 539  & 373 & 21 & 12 & 1562  \\
(d) sample with $z$ ($\geq85$\% CL) & 616 & 538  & 369 & 19 & 11 & 1553 \\
(e) sample used in MDLF             & 597 & 518  & 350 & (10) & (7) & 1482  \\
%(f) sample used to give $\phi^*$    & 367 & 297  & 234 & (5)  &     & 861 \\
(f) sample used to give $\phi^*$    & 314 & 368  & 253 & (5)  &     & 894 \\
\tableline
\end{tabular}
\end{center}
\end{table*}%

\clearpage

%table 2
\begin{table*}
\begin{center}
\caption{Luminosity function parameters}
\begin{tabular}{lcccc}
\tableline \tableline
morphology & $M^*({\rm r^{\ast}})-5\log h$ & $\alpha$ & $\phi ^\ast$ &  
 $\cal L({\rm r^{\ast}})$\\
           &                    &          &  $(0.01h^3 $Mpc$^{-3}$) &
 ($10^8h{\rm L}_\odot$ Mpc$^{-3}$)\\
\tableline
total            & $-20.65\pm 0.12$ & $-1.10\pm 0.14$  & $1.43\pm0.21$  & 2.00\\
$0\le T\le 1.0$  & $-20.75\pm 0.17$ & $-0.83\pm 0.26$  & $0.47\pm0.09$ & 0.62 \\
$1.5\le T\le 3 $ & $-20.30\pm 0.19$ & $-1.15\pm 0.26$  & $0.95\pm0.15$  & 1.00\\
$3.5\le T\le 5 $ & $-20.30\pm 0.20$ & $-0.71\pm 0.26$  & $0.43\pm0.05$  & 0.37 \\
$5.5\le T\le 6 $ & $\sim -20.0$     & $-1.9$           & $\sim0.04$ &   \\
\tableline
$C<0.35$ & $-20.62\pm 0.14$ & $-0.68\pm0.23$ & $0.67\pm0.12$ & 0.76 \\
$C>0.35$ & $-20.35\pm 0.19$ & $-1.12\pm0.18$ & $1.09\pm0.14$ & 1.17 \\
\tableline
\end{tabular} 
\end{center}
\end{table*}%


\clearpage

\begin{references}

\reference{} Abraham, R.~G., Valdes, F., Yee, H.~K.~C., \& van den Bergh, S.\ 1994, \apj, 432, 75 

\reference{} Bernardi et al. 2002a, to be published in AJ

\reference{} Bernardi et al. 2002b, to be published in AJ

\reference{} Binggeli, B., Sandage, A. \& Tammann, G. A. 1988, ARAA,
26, 509

\reference{} Blanton, M. et al. 2001, AJ, 121, 2358 

\reference{} Bromley, B. C. 1998, ApJ, 505, 25

\reference{} Doi, M., Fukugita, M., \& Okamura, S. 1993, MNRAS, 264, 832

\reference{} Dressler, A. \& Sandage, A. 1983, ApJ, 265, 664

\reference{} Efstathiou, G., Ellis, R. S. \& Peterson, B. A. 1988,
MNRAS, 232, 431

\reference{} Folkes, S. et al. 1999, MNRAS, 308, 459

\reference{} Fukugita, M., Ichikawa, T., Gunn, J. E., Doi, M., 
Shimasaku, K., \& Schneider, D. P. 1996, AJ, 111, 1748

\reference{} Fukugita, M., Shimasaku, K., \& Ichikawa, T. 1995,
PASP, 107, 945

\reference{} Fukugita, M. \& Turner, E. L. 1991, MNRAS, 253, 99

\reference{} Gunn, J. E. et al. 1998, AJ, 116, 3040

\reference{} Hogg, D. W., Finkbeiner, D. P., Schlegel, D. J., and Gunn,
 J. E. 2001, AJ, 122, 2129

\reference{} Kochanek, C. S. et al. 2001, ApJ, 560, 566 

\reference{} Kormendy, J. 1986, in Nearly Normal Galaxies, ed. S.
M. Faber (New York: Springer-Verlag), p. 163

\reference{} Loveday, J., Peterson, B. A., Efstathiou, G \& 
Maddox, S. J. 1992, ApJ, 390, 338 

%\reference{} Loveday, J., Tresse, L. \& Maddox, S. 1999, MNRAS, 310, 281

\reference{} Marzke, R. O., Geller, M. J., Huchra, J. P. \& Corwin, Jr.
H. G. 1994, AJ, 108, 437

\reference{} Marzke, R.~O., da Costa, L.~N., Pellegrini, P.~S., Willmer, C.~N.~A., \& Geller, M.~J.\ 1998, \apj, 503, 617 

\reference{} Meisels, A. \& Ostriker, J. P. 1984, AJ, 89, 1451

\reference{} Pier, J. R. et al. 2002, submitted to AJ

\reference{} Sandage, A. 1961, The Hubble Atlas of Galaxies.
Carnegie Institution of Washington, Washington DC

\reference{} Sandage, A., Tammann, G. A. \& Yahil, A. 1979, ApJ, 232, 
352

\reference{} Schlegel, D. J., Finkbeiner, D. P. \& Davis, M. 1998, 
ApJ, 500, 525

\reference{} Shimasaku, K. et al. 2001, AJ, 122, 1238 

\reference{} Smith, J. A. et al. 2002, ApJ, 123, 2121 

\reference{} Stoughton, C. et al. 2002, AJ, 123, 485

\reference{} Strateva, I. et al. 2001, AJ, 122, 1861 

\reference{} Strauss, M. A. et al. 2002, AJ, 124, 1810

\reference{} Tammann, G. A., Yahil, A. \& Sandage, A. 1979, ApJ, 234,
775

\reference{} de Vaucouleurs, G., de Vaucouleurs, A., Corwin, H., 
Buta, R., Paturel, G., \& Fouqu\'e, P. 1991, 
Third Reference Catalogue of Bright Galaxies (Springer, New York) (RC3)


\reference{} Yasuda, N. et al. 2001, AJ, 122, 1104

\reference{} York, D. G. et al. 2000, AJ, 120, 1579

\reference{} Zucca, E., Pozzetti, L. and Zamorani, G. 1994, MNRAS, 269, 953
\end{references}
\end{document}